\documentclass[prb,aps,superscriptaddress,reprint]{revtex4-1}

\usepackage{graphicx}
\usepackage[english]{babel}
\usepackage{SIunits}
\usepackage{color}

\begin{document}

\title{How the dynamic of photo-induced gate screening complicates the investigation of valley physics in 2D materials}

  \author{Frank Volmer}
   \affiliation{2nd Institute of Physics and JARA-FIT, RWTH Aachen University, 52074 Aachen, Germany}
  \author{Manfred Ersfeld}
   \affiliation{2nd Institute of Physics and JARA-FIT, RWTH Aachen University, 52074 Aachen, Germany}
  \author{Lars Rathmann}
 \affiliation{2nd Institute of Physics and JARA-FIT, RWTH Aachen University, 52074 Aachen, Germany}
  \author{Maximilian Heithoff}
   \affiliation{2nd Institute of Physics and JARA-FIT, RWTH Aachen University, 52074 Aachen, Germany}
  \author{Luca Kotewitz}
   \affiliation{2nd Institute of Physics and JARA-FIT, RWTH Aachen University, 52074 Aachen, Germany}
  \author{Kenji Watanabe}
  \affiliation{Research Center for Functional Materials, National Institute for Materials Science, 1-1 Namiki Tsukuba, Ibaraki 305-0044, Japan}
  \author{Takashi Taniguchi}
  \affiliation{International Center for Materials Nanoarchitectonics,National Institute for Materials Science, 1-1 Namiki Tsukuba, Ibaraki 305-0044, Japan}
  \author{Christoph Stampfer}
   \affiliation{2nd Institute of Physics and JARA-FIT, RWTH Aachen University, 52074 Aachen, Germany}
   \affiliation{Peter Gr\"unberg Institute (PGI-9), Forschungszentrum J\"ulich, 52425 J\"ulich, Germany}
  \author{Bernd Beschoten}
  \affiliation{2nd Institute of Physics and JARA-FIT, RWTH Aachen University, 52074 Aachen, Germany}
  \email{bernd.beschoten@physik.rwth-aachen.de}

\begin{abstract}
{An in-depth analysis of valley physics in 2D materials like transition metal dichalcogenides requires the measurement of many material properties as a function of Fermi level position within the electronic band structure. This is normally done by changing the charge carrier density of the 2D material via the gate electric field effect. Here, we show that a comparison of gate-dependent measurements, which were acquired under different measurement conditions can encounter significant problems due to the temporal evolution of the charging of trap states inside the dielectric layer or at its interfaces. The impact of, e.g., the gate sweep direction and the sweep rate on the overall gate dependence gets especially prominent in optical measurements due to photo-excitation of donor and acceptor states. Under such conditions the same nominal gate-voltage may lead to different gate-induced charge carrier densities and, hence, Fermi level positions. We demonstrate that a current flow from or even through the dielectric layer via leakage currents can significantly diminish the gate tunability in optical measurements of 2D materials.}
\end{abstract}
\maketitle

\section{Introduction}

One important goal in the field of 2D materials is the investigation of valley physics in semiconducting transition metal dichalcogenides (TMDs)~\cite{NatureReviewsMaterials.1.16055,PhysRevLett.108.196802}. As valley dynamics are governed by a delicate interplay of different electron-electron, electron-phonon and many-body interactions, an overall understanding of valley physics is only possible and physical models can only be tested when different device properties such as valley lifetimes, exciton lifetimes, spin and momentum scattering times, or phonon and electron dispersion relations are analyzed as a function of Fermi level position - ideally for the same device. To accomplish this, a variety of different measurements are required, most importantly the combination of both optical and electrical techniques. E.g. only the combination of gate-dependent electrical transport measurements, photoluminescence (PL) spectroscopy, and time-resolved Kerr rotation (TRKR) measurements recently enabled us to identify the dynamics which are responsible for the transfer of a polarization from optically excited bright trions to a valley polarization of free charge carriers in monolayer WSe$_2$ \cite{NanoLetters.20.3147}.

\begin{figure*}[tb]
	\includegraphics[width=\linewidth]{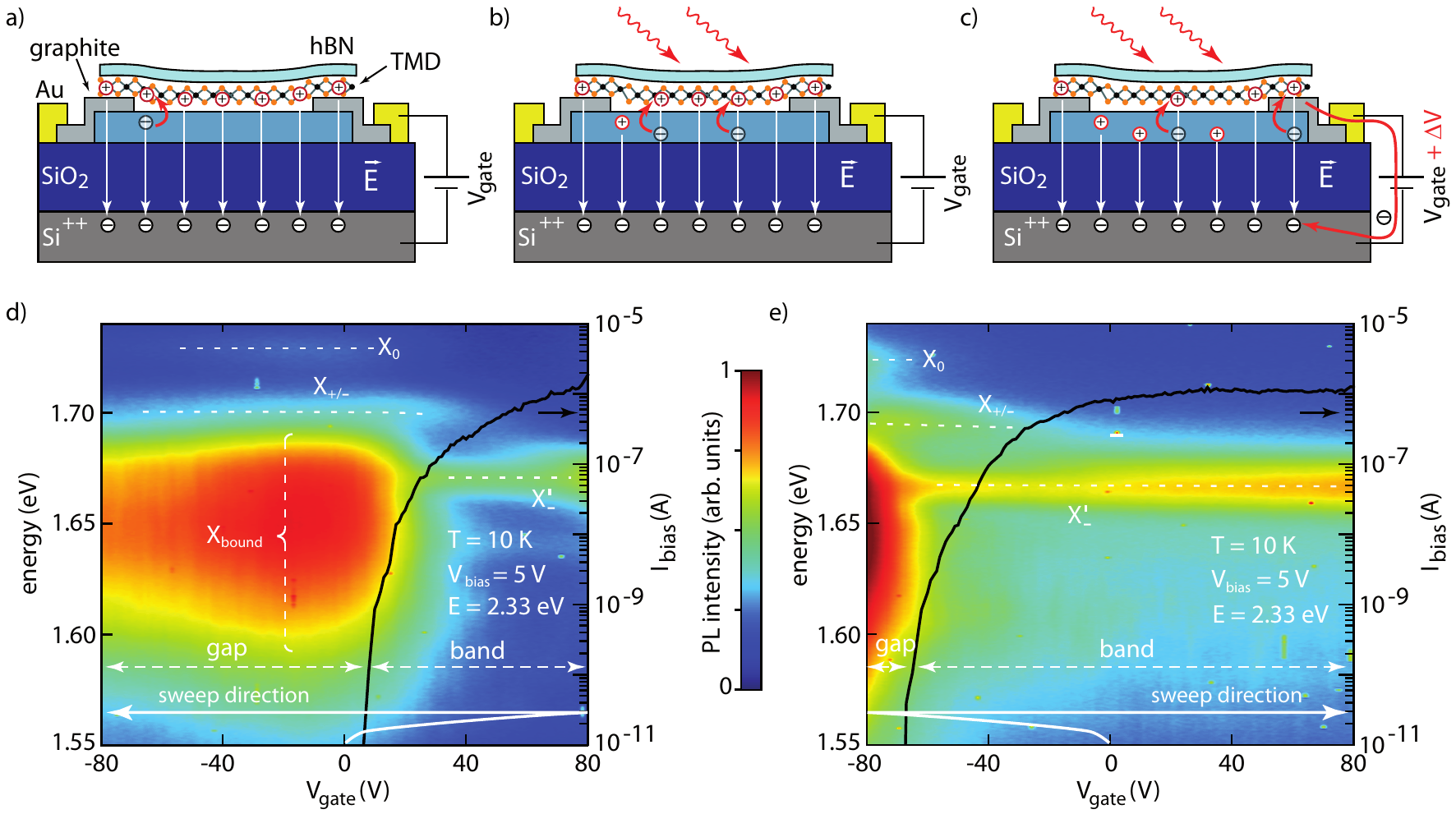}
	\label{fig1}
	\caption{(a) to (c): Schematic representation of the photo-induced gate-screening in case of donor states in the dielectric hBN layer. (a) Electrons from donor states can follow the gate electric field and tunnel into the 2D material, leaving behind a positively ionized trap state, which in turn screens the gate electric field (b). This process intensifies under illumination as more strongly bound donor states can now be excited. (c) The electrons from donor states will not lead to a change in the charge carrier density of the 2D material, if a simultaneous change in the gate voltage is transferring these electrons to the gate electrode. (d) and (e): Gate-dependent photoluminescence measurements of a monolayer WSe$_2$ device. Depending on the sweep direction of the gate voltage, the onset of the conduction band, seen in the exponential increase in the simultaneously measured electrical transfer characteristic, shifts significantly. A clear gate tunability is only achieved in the one half of the gate voltage axis, in which the gate sweep starts, whereas the Fermi level is almost pinned in the other half, as seen in the almost absent changes in both the bias current and the position and amplitude of excitonic features.}
\end{figure*}

However, this necessary prerequisite for the investigation of valley physics encounters practical obstacles, as not every measurement system can simultaneously conduct optical and electrical measurements. Even worse is the fact that some measurement techniques are mutually exclusive. E.g. in valley sensitive TRKR measurements the pump and probe energies normally have to be set to neutral or charged exciton energies (i.e. energies below the band gap) \cite{NanoLetters.20.3147,NatPhys.11.830,NanoLetters.19.4083}, but for this, first PL measurements with above-band-gap excitation are necessary to determine the energetic position of these exciton features \cite{RevModPhys.90.021001,PhysRevB.95.235408}. Therefore, it is important to compare gate-dependent measurements recorded under different conditions.

Here, we show that comparing gate-dependent data from different measurement techniques may easily lead to erroneous conclusions, as the same nominal gate-voltage may lead to different gate-induced charge carrier densities and, hence, Fermi level positions, depending on the measurement technique. This is due to donor and acceptor states created by impurities in the hexagonal boron nitride (hBN) layer which is used as a gate dielectric \cite{JournalofCrystalGrowth.303.525,Int.J.Appl.Ceram.Technol.8.977}. It is generally accepted that an electron transfer via these states leads to a charging of trap states (see red arrow in Figure~1a), although it is still under debate if these trap states are located either in the bulk of the hBN layer, the hBN-to-2D-material interface, or even the hBN-to-substrate interface~\cite{NatureNanotechnology.9.348,Nanoscale.11.7358,ACSAppl.Mater.Interfaces.8.9377,ACSApplMaterInterfaces.11.12170,2DMaterials.6.025040}. Independent on their exact position, these charged defects partially screen the gate-electric field (see the result in Figure~1b), which in turn will change the effective gate-induced charge carrier density in the 2D material.

This mechanism gets significantly more important as soon as the device is illuminated, as photo excitation of energetically strongly bound defect states becomes feasible (see additional charge transfer into the TMD by the optical excitation in Figure~1b) \cite{NatureNanotechnology.9.348,Nanoscale.11.7358,ACSAppl.Mater.Interfaces.8.9377}. It is important to note that such photo-induced changes in the electrical properties of a TMD device do not occur instantaneously but typically follow an exponentially decaying temporal evolution with significantly different time-scales of up to several minutes~\cite{NanoLetters.14.6165,AdvancedMaterials.29.1605598,Nat.Commun.10.4133}. This prevents a direct assignment of a gate voltage to a charge carrier density: Once the gate voltage is set to a fixed value, the time-dependent photo-induced screening of the gate-electric field will lead to a change in the gate-induced charge carrier densities and, therefore, a shift in the position of the Fermi level over time.

\section{Diminished gate tunability due to charging effects.}

An effective gate tuning of the 2D material, i.e. a change of its charge carrier density, can only be achieved as long as more charges per unit time are transferred via the voltage supply of the gate into the 2D material (see Figure~1c) than charges going either into defect states or contributing to a leakage current over the gate dielectric layer. The latter will be discussed in section~3. The rate at which charges are put into the 2D material via the voltage supply is directly proportional to the sweep rate of the gate voltage. For an ideal plate capacitor without any leakage current this charging current is given by:
\begin{equation}
    I=\frac{\mathrm{d}Q}{\mathrm{d}t}=\epsilon_0 \epsilon_r \frac{A}{d} \cdot \frac{\mathrm{d}V_\mathrm{gate}}{\mathrm{d}t},
\end{equation}
with the thickness $d$ of the dielectric layer, its dielectric constant $\epsilon_r$, the area $A$ of the 2D material, and the gate voltage $V_\mathrm{gate}$. On the other hand, the rate at which charges get transferred into defect states was shown to depend on both the photon energy \cite{NatureNanotechnology.9.348,Nanoscale.11.7358} and intensity \cite{ACSApplMaterInterfaces.11.12170}. Accordingly, we typically observe that a high light intensity or a slow gate sweep rate significantly diminishes the gate tunability of monolayer TMDs in PL measurements, similar to what is seen in electrical measurements \cite{ACSNano.6.5635}.

\begin{figure*}[tb]
	\includegraphics[width=\linewidth]{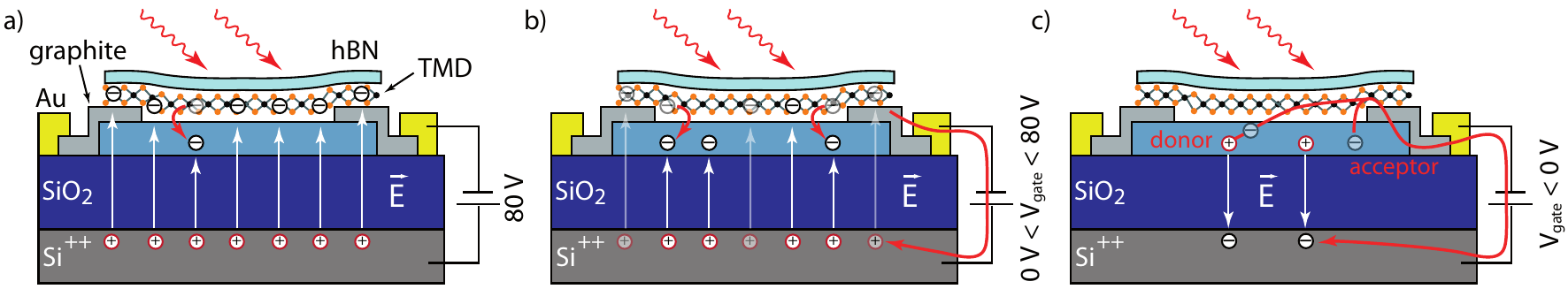}
	\label{fig2}
	\caption{Illustration of the diminished gate tunability seen for $V_\textrm{gate}<\unit[0]{V}$ of the gate-voltage sweep in Figure~1d. (a) Due to the initial fast gate sweep from $\unit[0]{V}$ to  $V_\textrm{gate}=\unit[+80]{V}$, the corresponding large flow of electrons into the 2D material made it highly n-doped. (b) In the gate voltage range from $\unit[+80]{V}$ to $\unit[0]{V}$ there are two ways of electron depletion within the TMD layer (see red arrows): 1.) The decreasing gate voltage leads to an electron flow over the voltage supply back to the gate-electrode (rightmost arrow). 2.) The continuous illumination with the excitation laser results in a transfer of electrons into acceptor states. These two currents  lead to an efficient tuning of the charge carrier density in the 2D material for decreasing positive gate voltages. (c) The efficient gate tuning stops as soon as negative gate voltages are applied which results in a reversal of the gate electric field direction. Electrons from previously ionised acceptor states start to flow back into the 2D material and so far neutral donor states also start to transfer electrons into the 2D material. The electron current from the dielectric layer into the TMD more or less compensates the electron current from the TMD over the gate voltage supply (see red arrows). As a result, there is no longer an effective change in the charge accumulation in the 2D material, i.e. almost no gate tunability.}
\end{figure*}

The PL measurements of Figures~1d and 1e demonstrate the impact of the sweep rate and also the sweep direction on the measured gate tunability of a monolayer WSe$_2$ flake. The monolayer TMD is sandwiched between two layers of hBN (thicknesses of around \unit[20-30]{nm}), placed onto a Si$^{++}$/SiO$_2$ (\unit[285]{nm}) wafer, and contacted via graphitic leads (see schematic in Figure~1a). These graphitic leads allow to simultaneously record the electrical transport in the TMD during PL measurements by applying a bias voltage $V_\textrm{bias}$ between the two contacts (black curves in Figures~1d and 1e show the corresponding gate-dependent bias currents). Due to these transport measurements we conclude that in Figure~1d the Fermi level is pinned inside the band gap of the TMD in the gate voltage range between \unit{-80}{V} and \unit{+10}{V}. This most likely results from mid-gap states, which are also responsible for the dominance of the bound exciton feature (X$_\text{bound}$) in the PL spectra \cite{RevModPhys.90.021001,PhysRevLett.121.057403,SciRep.3.2657}. Around \unit[+10]{V} the bias current exponentially increases as the Fermi level moves through the tail states of the conduction band \cite{NatureCommunications.5.3087,IEEEElectronDeviceLetters.39.761}. Above $V_\text{BG}=\unit[+20]{V}$ the Fermi level is in the conduction band, which results in a linear gate voltage dependence of the current and the disappearance of both the neutral (X$_0$) and bound exciton features.

The reversal of the sweep direction (compare white solid arrows in Figures~1d and 1e) now results in a significant shift of the gate voltage at which the Fermi level is tuned from the band gap into the conduction band (from around $\unit[10]{V}$ in Figure~1d to around $\unit[-65]{V}$ in Figure~1e). But the measurements do not just show a simple hysteresis effect, as in Figure 1e there is almost no gate tunability at all for higher positive gate voltages: The current becomes nearly constant, which is completely unexpected for an increasing number of free charge carriers at increasing gate voltages. We note that only at very high charge carrier densities, which are achievable in liquid-gate experiments, a saturation and eventual a dip in the conductance can be observed as soon as the Fermi level reaches the Q-valleys \cite{NanoLett.19.8836}. Furthermore, although the $X_-'$ feature in the PL measurement exhibits a small increase in amplitude, it nevertheless does not undergo the gradual redshift normally seen towards higher charge carrier densities \cite{NanoLett.17.740,NatureNanotechnology.12.144}.

A similar apparent lack of gate tunability can also be observed in Figure 1d, but at the opposite gate voltage range: Coming from positive gate voltages, the Fermi level seems to be pinned inside the band gap of the TMD over the whole negative gate voltage range. Overall, it seems that a clear gate tunability is only achieved in the one half of the gate voltage axis, in which the measurement starts, whereas it is pinned in the other one. To understand this observation, it is important to note that both measurements in Figures~1d and 1e start at a gate voltage of $V_\textrm{gate}=\unit[0]{V}$ with the excitation laser ($E=\unit[2.33]{eV}$) already illuminating the sample. Then the gate voltage is quickly swept with a rate of $\unit[4]{V s^{-1}}$ to either $V_\textrm{gate}=\unit[+80]{V}$ in Figure~1d or to $\unit[-80]{V}$ in Figure~1e, respectively. From there, the gate is swept towards the opposite gate voltage with a much lower average rate of around $\unit[0.4]{V s^{-1}}$ which is due to the necessary acquisition time of the PL spectra (see white solid arrows in Figures~1d and 1e).

We now discuss in detail the situation in Figure~1d: Before recording the very first PL spectrum at $V_\textrm{gate}=\unit[+80]{V}$, the fast gate sweep, which started at $\unit[0]{V}$, led to large flow of electrons from the gate electrode (which is now positively charged) over the voltage supply to the 2D material, which in turn is now n-doped (see Figure~2a). Already during this initial gate sweep, some of the electrons went into now negatively charged trap states (see transition marked by the red arrow in Figure~2a). For the following PL spectra, which were recorded in the gate voltage range from $\unit[+80]{V}$ to $\unit[0]{V}$, two mechanisms of electron depletion of the TMD layer have to be considered (see red arrows in Figure~2b): On the one hand, the decreasing gate voltage will lead to an electron flow over the voltage supply back to the gate-electrode (rightmost arrow). On the other hand, the continuous illumination with the excitation laser will lead to a transfer of electrons into acceptor states. Overall, these two different electron currents away from the TMD layer will lead to an efficient tuning of the charge carrier density in the 2D material for decreasing positive gate voltages.

The efficient gate tuning stops as soon as negative gate voltages are applied which results in a reversal of the gate electric field direction (see Figure~2c). On the one hand, electrons from previously ionised acceptor states now start to follow this reversed electric field direction and flow back into the 2D material. On the other hand, so far neutral donor states also start to transfer electrons into the 2D material. This leaves behind positively charged donor states which now screen the gate electric field (see Figure~2c). In the PL measurements of Figure~1d the gate sweep rate and the intensity of the excitation laser were set to such values that the electron current from the dielectric layer into the TMD more or less compensates the electron current from the TMD over the gate voltage supply (see red arrows in Figure~2c). As a result, there is no longer an effective change in the charge accumulation in the 2D material, i.e. almost no gate tunability. The same argumentation, but with reversed roles of acceptor and donor states, can be made for the PL measurement with reversed sweep direction shown in Figure~1e and, therefore, explains the almost lack of gate tunability for, in this case, positive gate voltages.

\begin{figure*}[tb]
	\includegraphics[width=\linewidth]{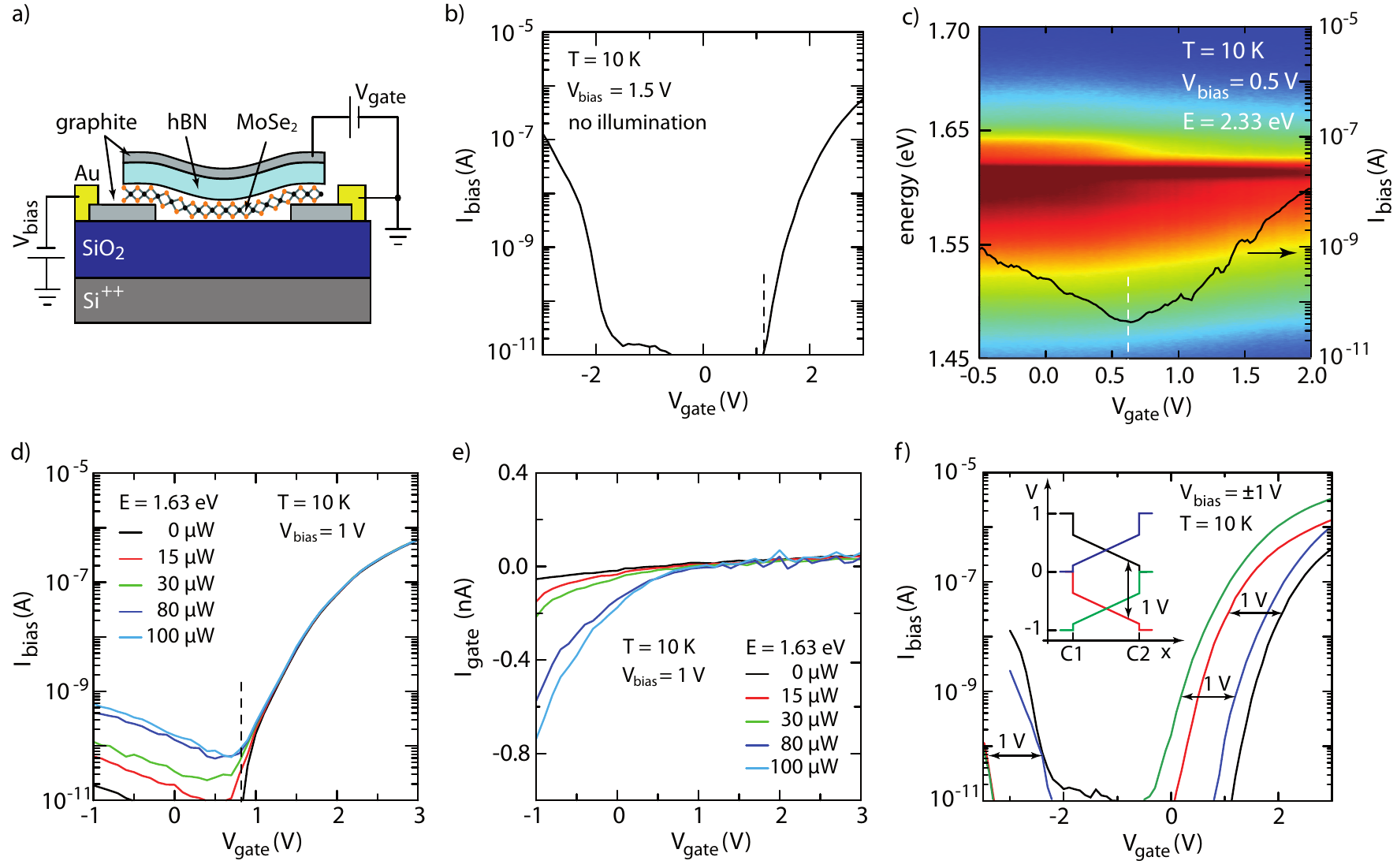}
	\label{fig3}
	\caption{(a) Schematic of a monolayer MoSe$_2$ device with a graphitic top gate. (b) Without  illumination, the gate-dependent electrical transport of this device shows a clear tuning of the Fermi level from the valence band over the band gap into the conduction band. (c) Instead, in the corresponding PL measurement both the exciton features and the bias current are smeared out. (d) For increasing laser powers an increase in sub-gap conductance can be observed. (e) The simultaneously measured leakage current over the gate dielectric shows a significant increase with laser power. Such a photo-induced leakage current creates a short between the graphitic gate and the TMD and therefore hinders the accumulation of charges. (f) As the gate electric field effect is based on the potential difference between gate electrode and 2D material, any bias voltage applied to the TMD can shift the gate-dependent charge transfer characteristic, as the gate voltage is normally referenced to the ground potential. Inset: Schematic voltage drop along the device in case that a bias voltage of $\unit[1]{V}$ is applied between two contacts, but for different current directions and grounding of either contact C1 or C2. The voltage drop at the contacts are due to Schottky barriers.}
\end{figure*}

\section{Diminished gate tunability due to leakage currents.}

In the following we discuss another current flow that can counteract an accumulation of charges in the 2D material. For this we investigate a device made from a monolayer MoSe$_2$ which is transferred onto a Si$^{++}$/SiO$_2$ (\unit[285]{nm}) wafer, contacted via graphitic leads, and gated via a graphitic top gate separated from the TMD by an hBN layer (see schematic in Figure~3a). Without illumination, the electrical transport between source and drain as a function of top-gate voltage shows a clear tuning of the Fermi level from the valence band over the band gap into the conduction band (see Figure~3b). Based on such a measurement we may expect clearly separated neutral and charged exciton features in a PL measurement. But instead, these features are completely smeared out in the actual PL measurement shown in Figure~3c. Furthermore, the simultaneously measured electrical transport (black curve in Figure 3c) no longer shows a clear suppression of conductance in the band gap region.

The appearance of a sub-gap conductance even with a laser excitation energy smaller than the band gap energy is demonstrated in Figure~3d as a function of laser power (the laser spot size had a full width at half maximum value of about $\unit[8]{\mu m}$ and the excitation energy was set to $\unit[1.63]{eV}$). Simultaneously to the measurement of the bias current $I_\mathrm{bias}$ shown in Figure~3d, also the current $I_\mathrm{gate}$ over the gate voltage supply was measured (see Figure~3e). In this measurement, a significant increase in leakage current with increasing laser power can be observed. Due to the wiring (see Figure~3a) this current has to flow over the top hBN layer. The appearance of such kind of photo-conductivity in hBN was previously attributed to stacking faults, nitrogen vacancies, or carbon and oxygen impurities \cite{AppliedSurfaceScience.138.364,PSSA.202.2229}. And even without illumination a charge transport in hBN was observed in the presence of structural defects or impurity atoms \cite{NanoLett.15.2263,ACSAppl.Mater.Interfaces.10.17287}. We note that the appearance and overall magnitude of such leakage currents significantly differ between different devices. It is the topic of ongoing investigations, if either the specific growth process, randomly distributed grain boundaries, or defects induced during the fabrication process, e.g. by reactive ion etching, are responsible for the observed variations in leakage currents. In any case, a leakage current over the dielectric layer as shown in Figure~3e shortens the two sides of the capacitor (i.e. the graphitic gate and the monolayer MoSe$_2$) and therefore hinders the accumulation of charges. This is a straightforward explanation for the vastly diminished gate electric field effect seen in the PL measurement of Figure~3c.

\section{Bias-dependent shift of charge transfer characteristic.}

In the previous sections we discussed that the assignment of a gate voltage to a charge carrier density is quite volatile in the presence of photo-induced effects. Now, we focus on the fact that even without light illumination it can be misleading if data from different measurements are plotted as a function of the applied gate voltage and then compared to each other. This can be clearly seen in the onsets of the conduction band marked by dashed lines in Figures~3b to 3d: In Figure~3b this onset is at a gate voltage slightly above $\unit[+1]{V}$, in Figure~3d it is slightly below $\unit[+1]{V}$, and finally in Figure~3c it is slightly above $\unit[+0.5]{V}$. This shift in gate voltage is not only due to the photo-induced gate-screening effect but also due to different bias voltages applied to the device, which were in the range of $\unit[+1.5]{V}$ to $\unit[+0.5]{V}$. For the gate electric field effect the important potential difference is not the one between gate electrode and the ground reference potential (i.e. $V_\textrm{gate}$ in the schematic of Figure~3a), but rather the potential difference between the gate electrode and the 2D material. In the presence of an additional bias voltage the potential of the 2D material is normally not equal to the ground potential. This is due to the formation of Schottky barriers at the interface between the contacting electrodes and the monolayer TMD \cite{NatMater.14.1195,NanoLett.13.3106} and the voltage drop across them. As in many experimental setups normally one contact is grounded, whereas the full bias voltage is applied to another contact (see $V_\textrm{bias}$ in the schematic of Figure~3a), the potential of the TMD lies somewhere between $V_\textrm{bias}$ and the ground reference potential, i.e. $\unit[0]{V}$.

Figure~3f demonstrates the impact of Schottky barriers on the gate-dependent electrical transport of the MoSe$_2$ device discussed so far. For these measurements, which were conducted without light illumination, always a voltage difference of $\unit[1]{V}$ is applied between two contacts labeled as C1 and C2 in the inset of Figure~3f. We switched both the contact which is put to the ground potential and the direction of the current by putting the contacts either to $\unit[+1]{V}$, $\unit[-1]{V}$, or $\unit[0]{V}$, respectively (each color represents one wiring configuration). One of the two Schottky contacts is always reverse-biased, whereas the other one is in forward bias direction. Therefore, the corresponding voltage drops at contacts C1 and C2 are different for the same wiring configuration and change with the current direction (compare the pair of black and red curves to the blue and green one). We furthermore assume a negligible voltage drop along the leads of the contacts and a constant voltage drop, i.e. constant resistivity, along the length of the TMD monolayer.

Overall, the only difference between the black and red curve (or the blue and green curve) is the shift of the potential at each point of the TMD flake by $\unit[1]{V}$ referenced to ground potential and, therefore, also to the gate voltage. As a result, there is exactly a $\unit[1]{V}$ shift seen in the gate-dependent electrical transport between these pairs of curves. Instead, the shift between e.g. the blue and black curve is far less straightforward as the highly non-linear I-V-characteristics of the Schottky contacts come into play as soon as the current direction is switched in the presence of two differently high Schottky barriers.

\section{Conclusion}

We have demonstrated that for optical measurements of 2D materials the assignment of a gate voltage to a charge carrier density can be extremely volatile: Once the gate voltage is set to a fixed value, the time-dependent photo-induced ionization of acceptor or donor states starts to screen the gate-electric field. This in turn leads to a change in the gate-induced charge carrier densities over time. The situation gets even more complicated if a gate sweep is conducted during the measurements. Under such conditions, an effective gate tuning of the 2D material, i.e. a change of the charge carrier density, can only be achieved as long as more charges per unit time are transferred via the voltage supply of the gate into the 2D material than charges going either into defect states or contributing to a leakage current over the gate dielectric layer. Overall, we demonstrate that great care has to be taken if gate-dependent measurements acquired under different measurement conditions are compared to each other that is, on the other hand, neccessary for an in-depth analysis of valley physics in 2D materials.

\section{Acknowledgement}

This project has received funding from the European Union's Horizon 2020 research and innovation programme under grant agreement No. 881603 (Graphene Flagship), the Deutsche Forschungsgemeinschaft (DFG, German Research Foundation)  under  Germany’s  Excellence  Strategy - Cluster of Excellence Matter and Light for Quantum Computing (ML4Q) EXC 2004/1 - 390534769 and by the Helmholtz Nanoelectronic Facility (HNF) at the Forschungszentrum J\"ulich \cite{HNF}. K.W. and T.T. acknowledge support from the Elemental Strategy Initiative conducted by the MEXT, Japan, Grant Number  JPMXP0112101001, JSPS KAKENHI Grant Number JP20H00354 and the CREST(JPMJCR15F3), JST.

\bibliographystyle{pss}
\providecommand{\latin}[1]{#1}
\makeatletter
\providecommand{\doi}
  {\begingroup\let\do\@makeother\dospecials
  \catcode`\{=1 \catcode`\}=2 \doi@aux}
\providecommand{\doi@aux}[1]{\endgroup\texttt{#1}}
\makeatother
\providecommand*\mcitethebibliography{\thebibliography}
\csname @ifundefined\endcsname{endmcitethebibliography}
  {\let\endmcitethebibliography\endthebibliography}{}

\end{document}